\documentclass{aa}
\usepackage{graphicx}
\usepackage[]{natbib}
\bibpunct{(}{)}{;}{a}{}{,} 
\usepackage{txfonts}
\newcommand{\kms}{~km\,s$^{-1}$}
\newcommand{\tk}{$T_\mathrm k$}
\newcommand{\n}{$n($H$_2$)}
\begin{document}
%
   \title{Sensitive CO and $^{13}$CO survey of water fountain stars.}
   \subtitle{Detections towards IRAS\,18460-0151 and IRAS\,18596+0315}

   \author{J. R. Rizzo\inst{1}
   \and
   J. F. G\'omez\inst{2}
   \and
   L. F. Miranda\inst{3,4}
   \and
   M. Osorio\inst{2}
   \and
   O. Su\'arez\inst{5}
   \and
   M. C. Dur\'an-Rojas\inst{2}}

   \institute{Centro de Astrobiolog\'{\i}a (INTA-CSIC),
              Ctra. M-108, km.~4, E-28850 Torrej\'on de Ardoz, Spain\\
              \email{ricardo@cab.inta-csic.es}
   \and
              Instituto de Astrof\'{\i}sica de Andaluc\'{\i}a (CSIC), 
              Apartado 3004, E-18080 Granada, Spain
   \and
              Consejo Superior de Investigaciones Cient\'{\i}ficas, Serrano 117,
              E-28006 Madrid, Spain
   \and
              Departamento de F\'{\i}sica Aplicada, Universidade de Vigo, 
              Campus Lagoas-Marcosende s/n, E-36310 Vigo, Spain
   \and
              UMR 6525 H.Fizeau, Universit\'e de Nice Sophia Antipolis, 
              CNRS, OCA. Parc Valrose, F-06108 Nice Cedex 2, France}
   
   \date{Received 1 July 2013; accepted 2013}

 
  \abstract
   {Water fountain stars represent a stage between the asymptotic giant branch 
    (AGB) and planetary nebulae phases, when the mass loss changes from 
    spherical to bipolar. These types of evolved objects are characterized by 
    high-velocity jets in the 22\,GHz water maser emission.}
   {The objective of this work is to detect and study in detail the circumstellar 
    gas in which the bipolar outflows are emerging. The detection and study of 
    thermal lines may help in understanding the nature and physics of the 
    envelopes in which the jets are developing.}
   {We surveyed the CO and $^{13}$CO line emission towards a sample of ten 
    water fountain stars through observing the $J=1\rightarrow0$ and 
    $2\rightarrow1$ lines of CO and $^{13}$CO, using the 30\,m IRAM radio 
    telescope at Pico Veleta. All the water fountains visible from the 
    observatory were surveyed.}
   {Most of the line emission arises from foreground or background Galactic 
    clouds, and we had to thoroughly analyse the spectra to unveil the velocity 
    components related to the stars. In two sources, IRAS\,18460-0151 and 
    IRAS\,18596+0315, we identified wide velocity components with a width of 
    $35-40$\,km\,s$^{-1}$ that are centred at the stellar velocities. These 
    wide components can be associated with the former AGB envelope of the 
    progenitor star. A third case, IRAS\,18286-0959, is reported as tentative; 
    in this case a pair of narrow velocity components, symmetrically located 
    with respect to the stellar velocity, has been discovered. We also modelled 
    the line emission using an LVG code and derived some global physical 
    parameters, which allowed us to discuss the possible origin of this gas in 
    relation to the known bipolar outflows. For IRAS\,18460-0151 and 
    IRAS\,18596+0315, we derived molecular masses close to 0.2\,M$_\odot$, mean 
    densities of $10^4$ cm$^{-3}$, and mass-loss rates of 
    $10^{-4}$\,M$_\odot$\,yr$^{-1}$. The kinetic temperatures are rather low, 
    between 10 and 50\,K in both cases, which suggests that the CO emission is 
    arising from the outer and cooler regions of the envelopes. No fitting was 
    possible for IRAS\,18286-0959, because line contamination can not be 
    discarded in this case.
   }
   {The molecular masses, mean densities, and mass-loss rates estimated for the 
    circumstellar material associated with IRAS\,18460-0151 and IRAS\,18596+0315 
    confirm that these sources are at the end of the AGB or the beginning of 
    the post-AGB evolutionary stages. The computed mass-loss rates are among the 
    highest ones possible according tocurrent evolutionary models, which leads us to 
    propose that the progenitors of these water fountains had masses in the range 
    from 4 to 8\,M$_\odot$. We speculate that CO emission is detected in water 
    fountains as a result of a CO abundance enhancement caused by current 
    episodes of low-collimation mass-loss. 
   }


   \maketitle
%

\section{Introduction}

Planetary nebulae (PNe) display a variety of morphologies, from purely 
spherical \citep[see, e.g.,][]{jac10} to bipolar or even multipolar 
\citep{sah98}. Non-spherical PNe are of particular interest, because the 
explanation of the process that shapes them is still under debate. There is 
growing evidence of highly collimated jets in evolved stars, which are believed 
to play a key role in the subsequent shaping of PNe \citep{sah98}. For low- and 
intermediate-mass stars ($0.8-8$ M$_\odot$), a short-lived transition from a 
spherical to a collimated mass-loss at some point between the asymptotic giant 
branch (AGB) and the PNe stages is therefore expected. Several processes 
have been proposed to explain the nature of collimated ejections in these 
evolved stars, such as the magnetic launching from a single star \citep{gar05} 
or a binary system \citep{sok98}. 

A crucial step in this study is identifying the sources that have just made the 
transition from spherical to collimated mass-loss. The group of objects often 
referred to as water fountain stars (hereafter WF) are probably the most 
appropriate candidates. WFs are evolved objects (mostly late AGB and post-AGB 
stars) with water maser emission tracing high-velocity motions, faster than the 
typical velocities seen in AGB mass-loss motions  of $10-30$\kms \citep[see, 
for example,][]{ner98}. Candidate objects to this class are usually identified 
by their broad velocity spread ($\ga 50$\kms) in their water maser spectra. 
Since the first identification of IRAS\,16342-3814 \citep{lik88}, this novel 
group of sources has grown very little in number. A total of only 14 candidate 
WFs have been reported so far \citep{eng86, dea07, deg07, sua08, sua09, day10, 
gom11}.

When water maser emission in WFs is observed with interferometers, it seems 
to trace highly collimated jets, with dynamical ages as young as $< 100$\,yr  
\citep{bob07, ima07a, day10}. In \object{IRAS\,18286-0959}, the water masers 
are tracing a double-helix structure \citep{yun11}. In all cases, these recent 
ejections represent one of the earliest known manifestations of collimated 
mass-loss in evolved stars.

WFs are thought to possess a thick, expanding circumstellar envelope (CSE), 
that was expelled during the AGB phase. The maser emission in these sources may 
be produced when a newly produced jet strikes into the CSE \citep{ima07a}. This 
scenario explains the highly collimated jets of OH and H$_2$O with velocities 
higher than 100\kms \citep{bob07, day10, gom11} and very short dynamical 
time-scales. After some decades, the tip of the jet is expected to have reached 
the outer regions of the CSE, where physical conditions for maser pumping are 
no longer met. The thick CSE in WF is evident by their high obscuration in 
optical wavelengths. 

The modelling of the broad-band spectral energy distribution (SED) of the 
thermal dust emission may give important clues about the envelope in WFs and 
also about the presence of disks (Dur\'an-Rojas et al., in preparation). SED 
modelling may, in principle, provide hints about the circumstellar mass, the 
luminosity of the system (e.g., including both the envelope and the disk), and 
the physical structure (density and temperature profiles). Unfortunately, SED 
modelling is still hampered by some free parameters that are poorly constrained 
due to the complexity of the sources (star, envelope, disks, jets) and the lack 
of knowledge of some global parameters.

The detection and detailed study of line emission from thermal gas is a key in 
providing additional valuable information about the physical characteristics of 
CSEs, mainly about the total mass, the mass-loss rate and the global kinematics.  
CO and $^{13}$CO spectra have been widely observed in AGB and post-AGB stars; 
modelling them has been used to determine the mass-loss rate, the total mass, and 
other physical parameters \citep[see, for example,][]{tey06}. So far, 
\object{IRAS\,16342-3814} is the only WF where thermal line emission has been 
unambiguously reported \citep{he08, ima09}. From the detection of the CO and 
(tentatively) $^{13}$CO $J=2\rightarrow1$ lines, \citet{he08} have determined 
an expansion velocity of 46 km\,s$^{-1}$, higher than that expected in an AGB 
envelope (typically 15\kms). Surprisingly, the CO $J=3\rightarrow2$ line 
reported recently \citep{ima12} shows a different kinematics, depicting an even 
broader profile ($>200$\kms). These high-velocity dispersions have been 
interpreted by the kinematics related to the collimated ejection also traced by 
water masers, instead of with the expanding motions of the CSE.

\citet{ima09} illustrated the difficulty of properly identifying CO associated 
with WFs when using single-dish observations. Most of these objects are at low 
Galactic latitude and therefore multiple Galactic foreground and background CO 
components are also gathered within the telescope beam. It is necessary to take 
spectra offset from the target position, and identify whether there are CO 
components that are present only at the target position, and at a velocity 
close to that of the star. \citet{ima09} reported a tentative detection of CO 
$J=3\rightarrow2$ emission associated with IRAS\,18286-0959, since it was only 
present in spectra taken towards this source. However, its large velocity offset 
($\simeq 25$\kms) from the central velocity of the jet and the narrow width of 
the line precluded these authors to ascertain the association of the CO 
emission with the WF.

We conducted sensitive CO and $^{13}$CO observations of ten WFs, involving the 
$J=1\rightarrow0$ and $2\rightarrow1$ transitions, using the 30m IRAM radio 
telescope at Pico Veleta. Our goal was to survey the largest possible number of 
WFs to obtain new reliable detections of thermal line emission associated with 
these objects, and hence to derive some global physical parameters of the 
emitting regions.

%
\begin{table*}[]
\caption{Sources observed}             
\label{table:1}      
\centering                          
\begin{tabular}{cccrcrrc}        
\hline\hline                 
IRAS name & Other name & RA & \multicolumn{1}{c}{Dec} & \multicolumn{1}{c}{$d$} & \multicolumn{1}{c}{$b$} & \multicolumn{1}{c}{$V_{\rm LSR}$} & Ref. \\
&& \multicolumn{2}{c}{J2000} & \multicolumn{1}{c}{kpc} & \multicolumn{1}{c}{deg} & \multicolumn{1}{c}{\kms} \\
\hline                        
{\bf 16552-3050} & GLMP\,498  & 16:58:27.30 & $-$30:55:08.0 & 19.6 & $+7.3$ & $+20$  & 1,2\\
{\bf 18043-2116} & OH0.9-0.4  & 18:07:20.86 & $-$21:16:10.9 &  6.4 & $-0.4$ & $+90$  & 3\\
{\bf 18113-2503} &            & 18:14:27.27 & $-$25:03:00.5 &  6.7 or 10.0 & $-3.6$ & $+100$ & 4 \\
18139-1816? & {\bf OH12.8-0.9} & 18:16:49.23 & $-$18:15:01.8 &  --- & $-0.9$ & $+60$  & 5,6\\
{\bf 18286-0959} &            & 18:31:22.95 & $-$09:57:21.1 &  3.6 & $-0.1$ & $+41$  & 7,8\\
18450-0148 & {\bf W43A}       & 18:47:41.16 & $-$01:45:11.5 &  2.6 & $0.0$  & $+34$  & 9,10\\
{\bf 18460-0151} & OH31.0-0.2 & 18:48:43.02 & $-$01:48:30.5 &  2.1 & $-0.2$ & $+125$ & 11,12\\
{\bf 18596+0315} & OH37.1-0.8 & 19:02:06.26 & $+$03:20:15.5 &  4.6 or 8.8 & $-0.8$ & $+90$  & 13\\
{\bf 19134+2131} &            & 19:15:35.22 & $+$21:36:33.9 &  8.0 & $+4.6$ & $-65$  & 14\\
{\bf 19190+1102} &            & 19:21:25.09 & $+$11:08:41.0 &  8.6 & $-1.5$ & $+20$  & 15\\

\hline                                   
\end{tabular}

\tablebib{(1)~\citet{sua07}; (2)~\citet{sua08}; (3)~\citet{dea07}; 
(4)~\citet{gom11}; (5)~\citet{eng86}; (6)~\citet{bob05}; (7)~\citet{ima13a}; 
(8)~\citet{yun11}; (9)~\citet{dia85}; (10)~\citet{ima05}; (11)~\citet{deg07}; 
(12)~\citet{ima13b}; (13)~\citet{gom94}; (14)~\citet{ima07b}; (15)~\citet{day10}.
}

\tablefoot{In boldface are given the most common name and that used throughout 
           this paper.\\
           Unclear association of IRAS\,18139-1816 to OH12.8-0.9, probably due 
           to wrong coordinates in the IRAS catalogue.\\
           Kinematic distances provided for IRAS\,18113-2503 (see text).\\
           Unknown distance to OH12.8-0.9. \citet{bau85} proposed 8\,kpc due to 
           its membership to a group of evolved stars located towards the 
           Galactic Center.
          }
\end{table*}
%

\section{Observations and strategy}

The observations were carried out using the 30\,m radio telescope at Pico 
Veleta, Spain, during two runs in June 2009 and June 2010. The new EMIR 
(Eight MIxer Receiver) was used for all the observations. The focal plane 
geometry of this instrument allowed us to gather 3\,mm and 1\,mm data 
simultaneously in two linear polarizations. In this case, we selected 
the simultaneous observations of CO $J=1\rightarrow0$ (115.271\,GHz) 
and $J=2\rightarrow1$ (230.538\,GHz) in one receiver configuration, and 
$^{13}$CO $J=1\rightarrow0$ (110.201\,GHz) and $J=2\rightarrow1$ 
(220.399\,GHz) in a second setup. 

Both the VESPA and WILMA autocorrelators were connected as backends at all the 
observed lines. With VESPA, a frequency spacing of 312\,kHz and 1.25\,MHz was 
employed in the 3\,mm and 1\,mm lines. This is equivalent to 0.81 and 0.85\kms\ 
at the CO and $^{13}$CO $J=1\rightarrow0$ frequencies; at the $J=2\rightarrow1$ 
lines, the velocity spacings are twice as high. The theoretical bandwidth 
employed was 160\,MHz and 320\,MHz at 3\,mm and 1\,mm, which provided an actual 
velocity coverage of $\sim 360$\kms\ in all the lines. On the other hand, each 
module of the WILMA autocorrelator provided 1\,GHz of bandwidth and a fixed 
spectral resolution of 2\,MHz. The wide instantaneous bandwidth of EMIR (8 GHz 
in SSB at 3\,mm) allowed us to tune the local oscillator to include the 
$^{13}$CO $J=1\rightarrow0$ line in the CO setup. Therefore, we used another 
WILMA module to cross-check the $^{13}$CO $J=1\rightarrow0$ line emission 
within the CO setup, albeit with a poor velocity resolution (5.4\kms).

The observations were made under typical winter conditions (precipitable 
water vapour column close to 4\,mm), registering atmospheric opacities at 
225\,GHz between 0.1 and 0.4. Typical system temperatures were 110--450\,K 
at 3\,mm, and 160--600\,K at 1\,mm. The resulting {\it rms} (three-sigma) were 
between 8\,mK and 92\,mK at 3\,mm, and between 26\,mK and 250\,mK at 1\,mm.

The antenna temperature scale was calibrated every 10--12 minutes by the 
standard chopper wheel method, that is, by the sequential observation of hot and 
cold loads, and the blank sky; hot and cold load temperatures were room- and 
liquid-nitrogen temperatures, respectively. Sky attenuation was determined 
in real time from the values of a weather station, the measurement of the sky 
emissivity, and an atmosphere numerical model available at the observatory. 
Furthermore, we regularly observed a set of standard line calibrators 
\citep{mau89}, and estimated an uncertainty lower than 20\% in all the 
observed lines.

All the intensities throughout the paper are made with respect to a scale of 
main-beam temperature ($T_{\mathrm{MB}}$), assuming main-beam efficiencies of 
0.75 and 0.52 at 3\,mm and 1\,mm. Velocities are given with respect to the LSR.

We surveyed all the WFs known in 2010 that are visible from Pico Veleta. The 
sample includes ten sources, detailed in Table 1. In the table we included 
other common names for the sources, their equatorial coordinates (J2000), 
distances, Galactic latitudes, and approximate velocities with respect to the 
LSR. The two distances quoted for IRAS\,18113-2503 are kinematic distances 
derived in this work, using the Galactic rotation model of \citet{rei09}.

All the observed sources are located in the Galactic disk, and most of them at 
Galactic latitudes as low as $|\,b\,|<1\degr$. Therefore, the CO and $^{13}$CO 
lines are expected to be heavily contaminated by Galactic background or 
foreground emission that originates in local gas and spiral arm clouds along 
the line of sight. 

Therefore, we selected position-switching as the observing mode and were 
especially careful in selecting an appropriate reference position. For each 
source, we tried different reference positions by moving away from the 
Galactic plane, until we obtained clean spectra, that is, without line emission 
arising from the reference position.

However, a position used for reference that is too distant usually produces a 
worsening in the baselines, and we had to compromise. Typical distances to the 
reference positions were 1000\arcsec.

To associate a velocity component to a WF, we proceeded with a strategy of 
five-points crosses, that is, by observing towards the star position, and 
another four points located $24\arcsec$ away from it, in the east-west and 
north-south directions. This separation of $24\arcsec$ corresponds 
approximately to the half-power beam width (HPBW) at 3\,mm, and twice the HPBW 
at 1\,mm. Due to galactic foreground/background gas clumps, one can expect 
significant variations in the emission along $24\arcsec$. However, if we saw a 
velocity component present ONLY at the star position, this was suspected to be 
associated to the WF, and was analysed in detail.

When a WF source had no CO emission at the star position or only a narrow 
component far away from the stellar velocity, it was excluded from the 
five-points strategy. Because of time restrictions, we were unable to observe 
the $^{13}$CO setup in \object{IRAS\,16552-3050} and \object{OH12.8-0.9}; in 
these sources, we were only able to use the $J=1\rightarrow0$ spectra from 
WILMA, which were observed simultaneously with the $^{12}$CO setup (as 
explained above).

Data reduction and analysis were made using the GILDAS 
software\footnote{GILDAS is a radio astronomy software developed by IRAM. See {\tt http://www.iram.fr/IRAMFR/GILDAS}/.}.

\section{Results}
\subsection{Overview}

In Figs.~1 to 4 the spectra towards the star position are depicted, 
corresponding to CO $J=1\rightarrow0$, $^{13}$CO $J=1\rightarrow0$, CO 
$J=2\rightarrow1$, and $^{13}$CO $J=2\rightarrow1$. In all cases but the 
$^{13}$CO $J=1\rightarrow0$ line of IRAS\,16552-3050 and OH12.8-0.9 (Fig.~2), 
spectra are taken from the VESPA autocorrelator; for non-detections, a 
smoothing of three channels was applied to the spectra. Line emission, with 
multiple velocity components, is present in most of the spectra, but it mainly 
arises in foreground and background Galactic clouds. Notable exceptions are 
IRAS\,16552-3050, \object{IRAS\,18113-2503} (both without CO emission), and 
\object{IRAS\,19134+2131} (with only a narrow component), which are the sources 
with the highest Galactic latitude in our sample. Table 2 provides the upper 
limits of the CO and $^{13}$CO velocity-integrated temperatures in these cases, 
assuming a line width of 40\kms.

%
\begin{table}[]
\caption{Upper limits of the undetected sources}
\label{table:2}      
\centering                          
\begin{tabular}{ccccc}        
\hline\hline                 
IRAS name & \multicolumn{2}{c}{CO}            & \multicolumn{2}{c}{$^{13}$CO}\\
          & $1\rightarrow0$ & $2\rightarrow1$ & $1\rightarrow0$ & $2\rightarrow1$\\
\hline                        
16552-3050 & 0.83 & 0.64 & 0.36 & ---  \\
18113-2503 & 1.08 & 2.15 & 0.25 & 0.27 \\
19134+2131 & 0.61 & 0.72 & 0.53 & 1.61 \\
\hline                                   
\end{tabular}

\tablefoot{Three-sigma {\it rms} limits in K\,\kms. Velocity width of 40\kms\ assumed.
          }
\end{table}
%

%

   \begin{figure*}
   \centering
   \includegraphics[angle=-90, width=18.2cm]{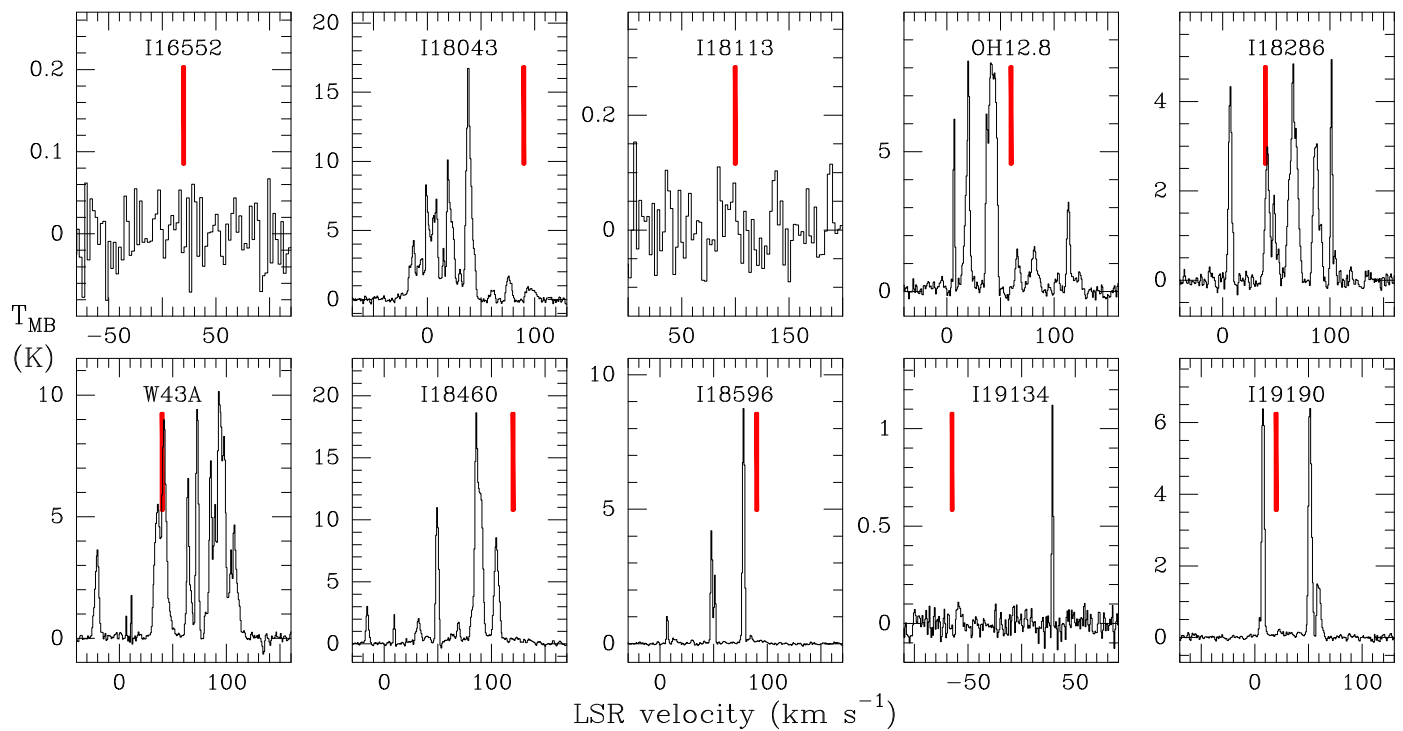}
   \caption{
CO $J=1\rightarrow0$ spectra towards the ten WFs surveyed. The (shortened) 
source name is indicated at the top of each spectrum. Note the different intensity 
ranges in the panels. All the spectra span 200\kms\ of coverage to facilitate 
line width comparisons among the sources. Spectra corresponding to 
IRAS\,16552-3050 and IRAS\,18113-2503 have been smoothed to three times the 
original velocity spacing. Red vertical bars are located at approximately the 
stellar velocity to facilitate additional association. Most of these velocity components 
arise from background or foreground Galactic clouds, as discussed in the text.
              }
   \end{figure*}
%

   \begin{figure*}
   \centering
   \includegraphics[angle=-90, width=18.2cm]{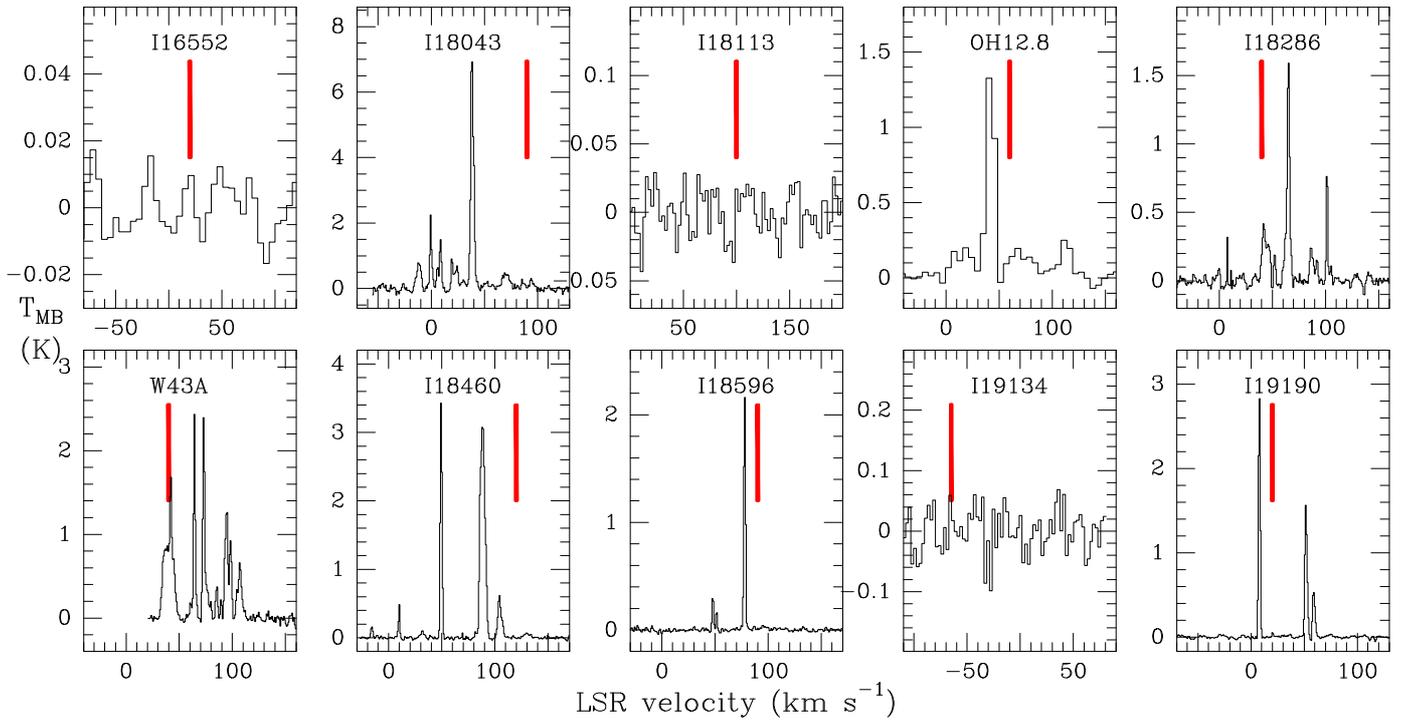}
   \caption{
Same as Fig.~1 for the $^{13}$CO $J=1\rightarrow0$ line. Spectra from 
IRAS\,16552-3050 and OH12.8-0.9 correspond to the WILMA autocorrelator. Spectra 
corresponding to IRAS\,18113-2503 and IRAS\,19134+2131 have been smoothed to three 
times the original velocity spacing.
             }
   \end{figure*}
%

   \begin{figure*}
   \centering
   \includegraphics[angle=-90, width=18.2cm]{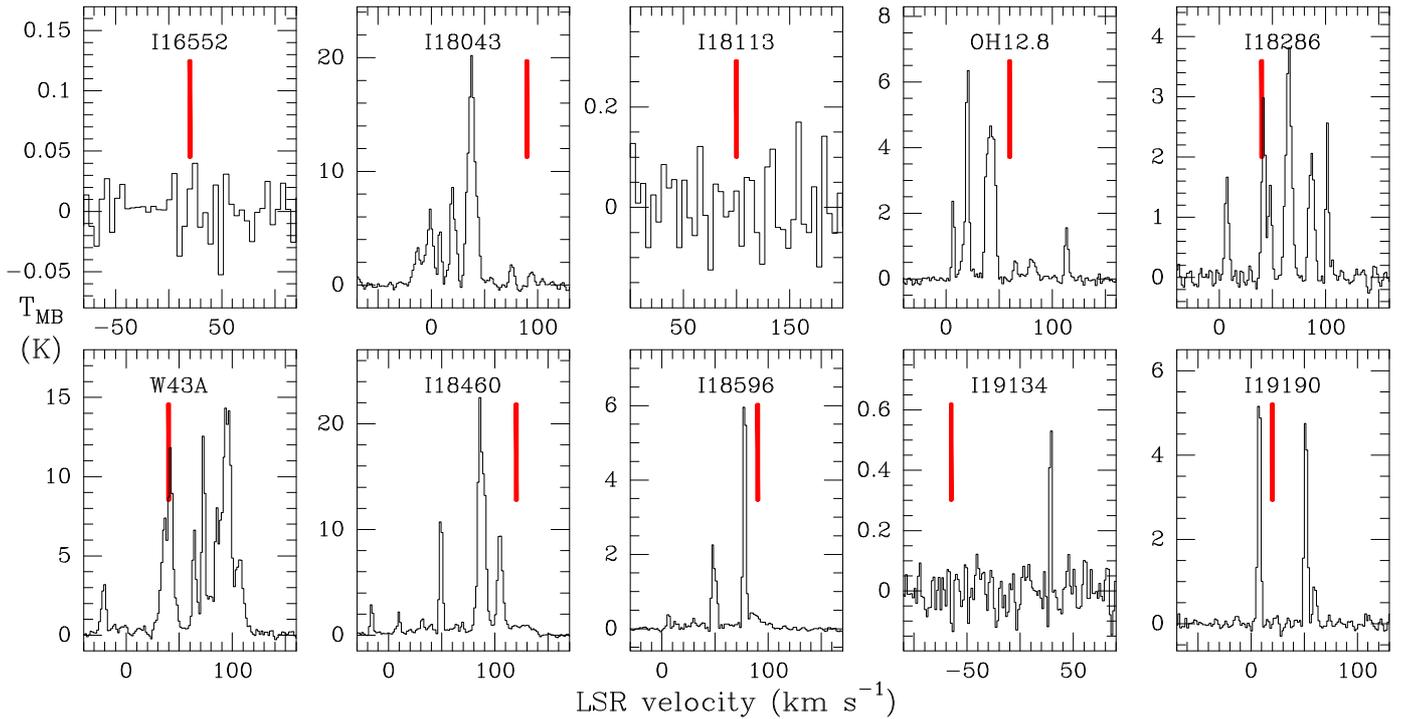}
   \caption{
Same as Figs.~1 and 2 for the CO $J=2\rightarrow1$ line. Spectra 
corresponding to IRAS\,16552-3050 and IRAS\,18113-2503 have been smoothed to three 
times the original velocity spacing.
              }
   \end{figure*}
%

   \begin{figure*}
   \centering
   \includegraphics[width=13.7cm]{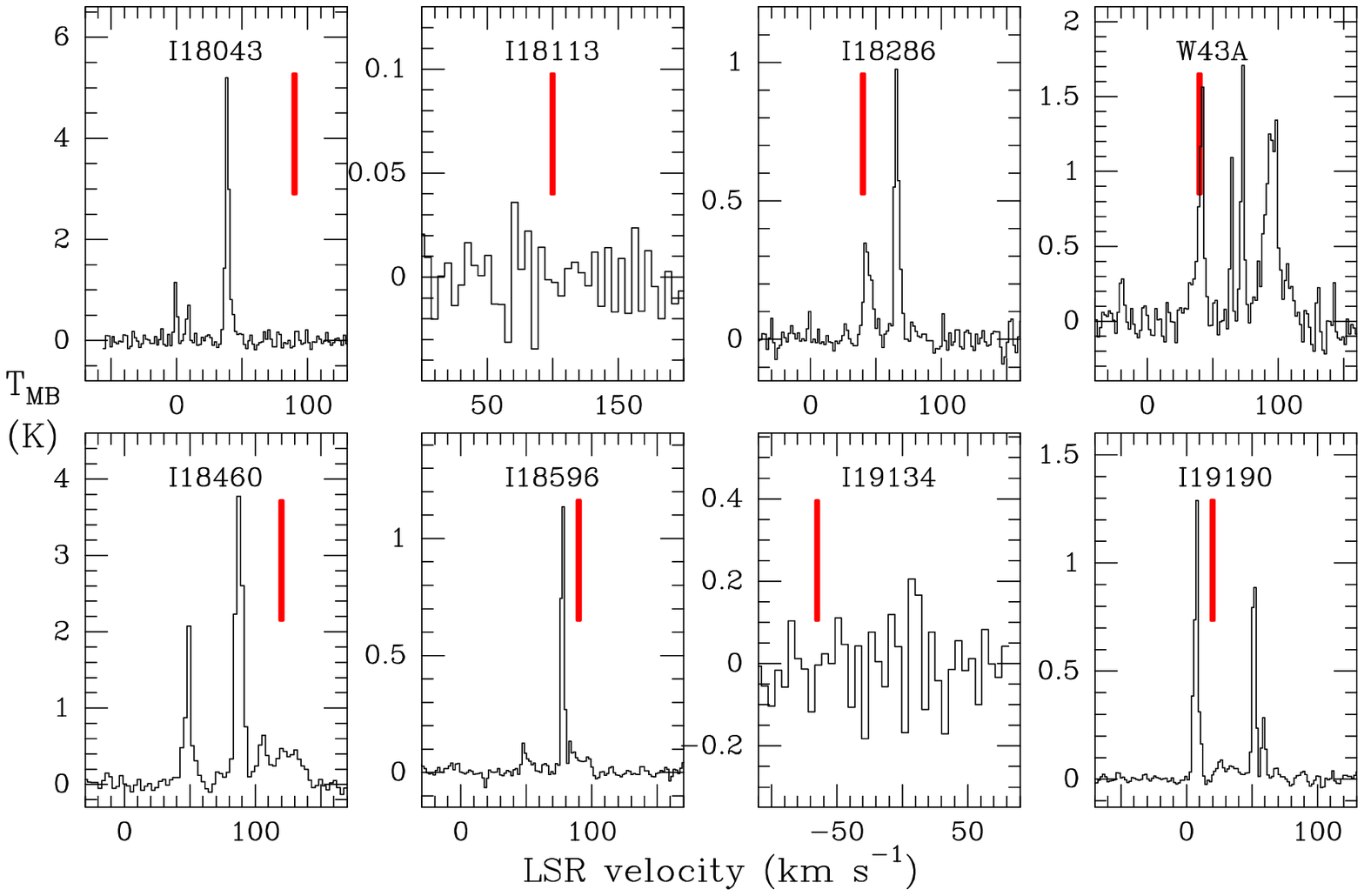}
      \caption{
Same as Figs.~1--3 for the $^{13}$CO $J=2\rightarrow1$ line. The sources 
IRAS\,16552-3050 and OH\,12.8-0.9 have not been observed in this transition. 
Spectra corresponding to IRAS\,18113-2503 and IRAS\,19134+2131 have been 
smoothed to three times the original velocity spacing.
              }
   \end{figure*}
%

The foreground and background molecular emission makes it difficult to ascertain 
the presence of emission associated with our target sources. Therefore, special 
care must be taken in finding this association. We propose a set of three 
requirements that a particular velocity component must fulfil for it to be 
considered as probably associated with the WFs: 
(1) it is present {\it only} at the star position; 
(2) it persists at least in two of the observed lines; and 
(3) it is centred as closely as possible at the stellar velocity.

These three criteria are independent of any additional interpretation of the 
geometry and kinematics of the CO-emitting region in WFs, and we consider them 
as indispensable requirements. In addition to these considerations, another 
reliability restriction can be imposed if we assume that CO emission arises in 
an expanding circumstellar shell, or in the region where this shell interacts 
with the stellar jet, or even in the inner disk. Therefore, we expect lines to 
have widths consistent with expanding motions found in AGB envelopes, or larger 
($\simeq 15$\kms). 

After a thorough analysis of the five-points pattern of all sources and all 
emission lines, we identified two strong candidates that might possess thermal gas 
associated with them: \object{IRAS\,18460-0151} and \object{IRAS\,18596+0315}; 
as we see in the next section, broad components are detected towards these 
sources. In a third case, \object{IRAS 18286-0959}, we discovered a pair of 
narrow CO velocity components, symmetrically located with respect to the mean 
OH and H$_2$O maser velocities \citep{sev97, ima13a}. While this source does 
not meet the three detection criteria we imposed, we cannot discard that the CO 
emission is associated with this WF, although with circumstellar 
characteristics different from those of the other two detections. We therefore 
include this source as a possible (albeit very tentative) detection in the 
following discussion.

\subsection{Individual sources}
\subsubsection{IRAS\,18460-0151}

This is the clearest detection reported in this work and, if confirmed, would 
be the second WF known to show associated thermal line emission, after 
IRAS\,16342-3814 \citep{he08, ima09, ima12}. Figure 5 depicts all the final 
spectra towards this source; for each of the observed lines, a cross of panels 
shows the spectra at the five positions observed. To facilitate the 
identification of the associated feature, a zoom in intensity and velocity was 
applied. 

%

   \begin{figure*}
   \centering
   \includegraphics[width=15cm]{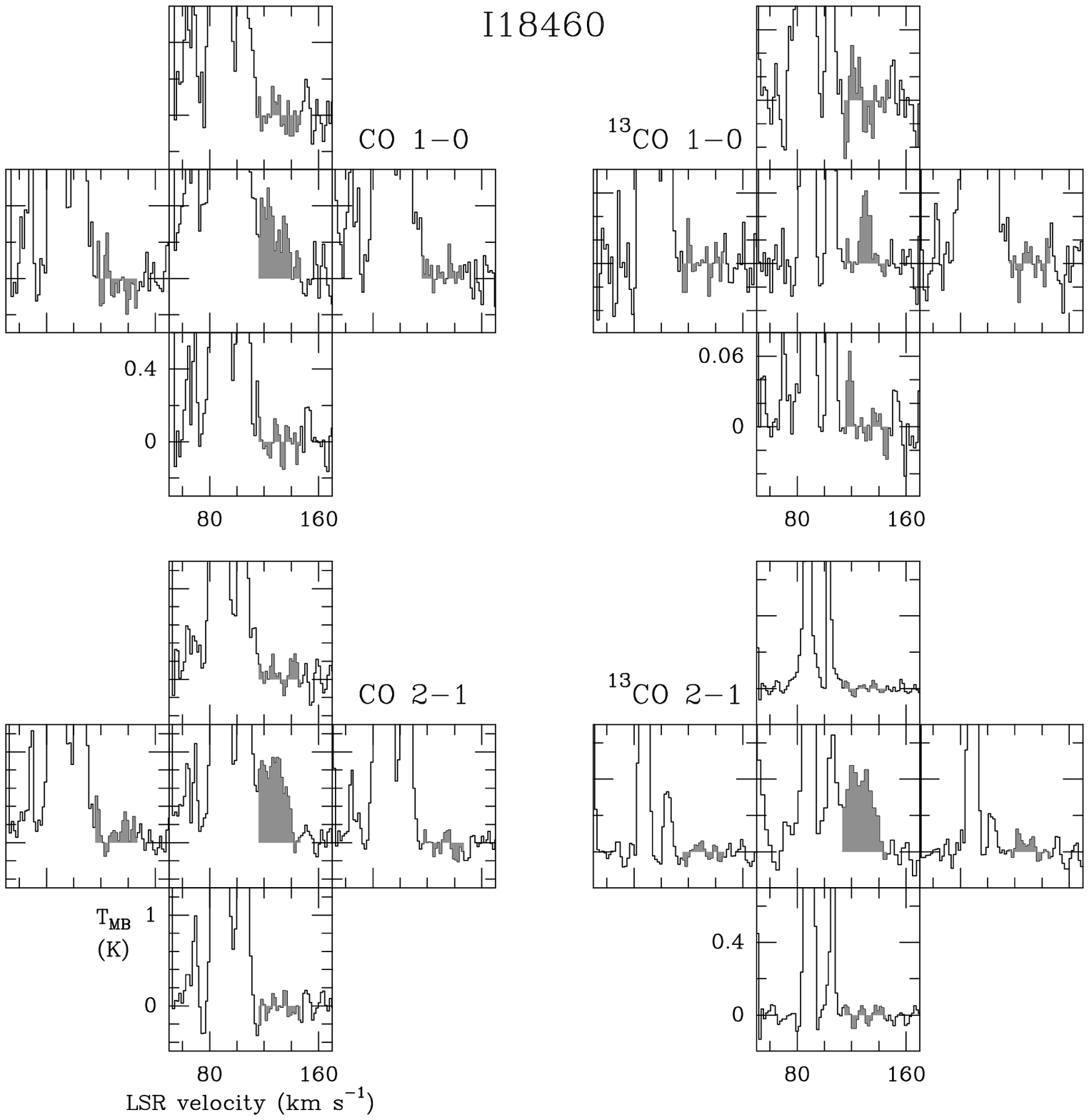}
      \caption{
All spectra observed towards IRAS\,18460+0151. Each cross of five spectra 
corresponds to one of the CO lines. The relative locations of the spectra 
indicate the relative positions of antenna pointings; the centre towards the WF 
and four offset positions, 24\arcsec away from the WF. The lines are indicated 
on top of each cross. The velocity range from 115 to 145\kms\ has been shaded 
to facilitate additional comparison. As discussed in the text, this velocity 
range depicts significant emission only at the star position.
              }
   \end{figure*}
%

A component of about 40 km\,s$^{-1}$ width was detected {\it in the four lines} 
observed, and {\it only} towards the star position. In addition, this component 
is centred close to the stellar velocity, inferred from the double-peaked OH 
spectrum \citep{tel89, eng07} and also the H$_2$O emission \citep{deg07}. To 
emphasize the feature, the velocity range $(115, 145)$\kms has been shaded 
in the figure. This is the only velocity range where the above mentioned 
characteristics are found for this source.

The contrast between the star and the off-source positions is more notable in 
the two $J=2\rightarrow1$ lines. In contrast, the $^{13}$CO $J=1\rightarrow0$ 
line shows a velocity component significantly narrower than the other lines. 

The line ratio of CO (2-1)/(1-0) is higher than 2. This is an unusually high 
value for extended insterstellar clouds \citep[see, e.g., ][]{sak95}, but it is 
more typical of a compact, unresolved source, which reinforces the association 
of this component to the CSE of the WF. The CO(2-1)/$^{13}$CO(2-1) ratio is 
remarkably low (between 2 and 3). This is also uncommon for an interstellar 
cloud, but similar values have been found in the WF IRAS\,16342-3814 
\citep{he08,ima12} and some AGB and post-AGB envelopes 
\citep[see, for example,][]{buj01,buj05,tey06}. This point is discussed in 
Sect.~4, where we analyse the emission and estimates of physical parameters.

\subsubsection{IRAS\,18596+0315}

This case is quite similar to IRAS\,18460-0151, although the associated CO 
and $^{13}$CO components are weaker than those of that source. Figure 6 
displays all the final spectra for this source, similar to the previous figure. 
Again, a zoom in intensity and velocity was applied to emphasize the 
feature of interest. 

%

   \begin{figure*}
   \centering
   \includegraphics[width=15cm]{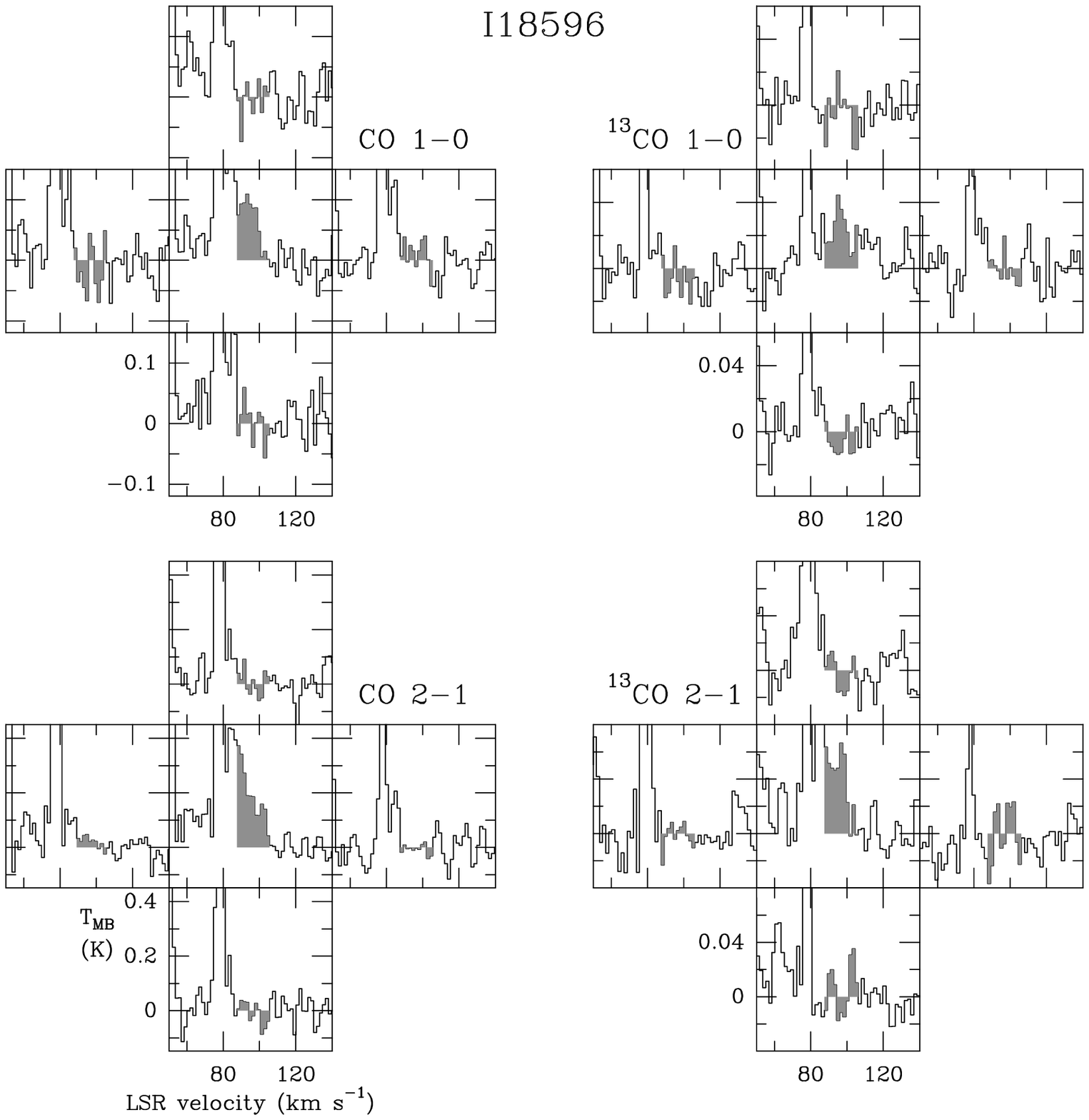}
      \caption{
Same as Fig.~5 for IRAS\,18596+0315. This is similar to IRAS\,18460+0151 and 
presents a wide component only at the WF position. The shaded area reaches from 
90 to 105\kms.
              }
   \end{figure*}
%

In the velocity range $(90, 105)$\kms\ (shaded in the figure), there is 
significant emission only towards the source, and none elsewhere. As in the 
previous case, this is noted in all the spectral lines observed in the survey. 
The whole range of emission of this feature is indeed wider, as we show in the 
discussion. As in IRAS\,18460-0151, the $^{13}$CO $J=1\rightarrow0$ line is the 
less evident, while the two $J=2\rightarrow1$ lines are the clearest. 

The line ratio of CO (2-1)/(1-0) is unusually high (around 2), as in 
IRAS\,18460-0151. The CO/$^{13}$CO observed ratios are not as low as in 
IRAS\,18460-0151; they vary between 3 and 5, depending on the 
$J\rightarrow(J-1)$ used for the computation. The implications of this 
difference is also discussed below.

\subsubsection{IRAS\,18286-0959}

This case is clearly different from the previous ones, because we did not find 
a wide velocity component. However, we did detect a pair of narrow lines, 
symmetrically located with respect to the stellar velocity. Figure 7 depicts 
all the final spectra towards this source, similar to Figs.~5 and 6. 
Unfortunately, we dealt with strong contamination in the offset positions of 
the two $^{13}$CO lines at one of the velocity ranges of interest. For this 
reason, Fig.~7 contains only the CO lines at the five positions observed.

%

   \begin{figure*}
   \centering
   \includegraphics[width=15cm]{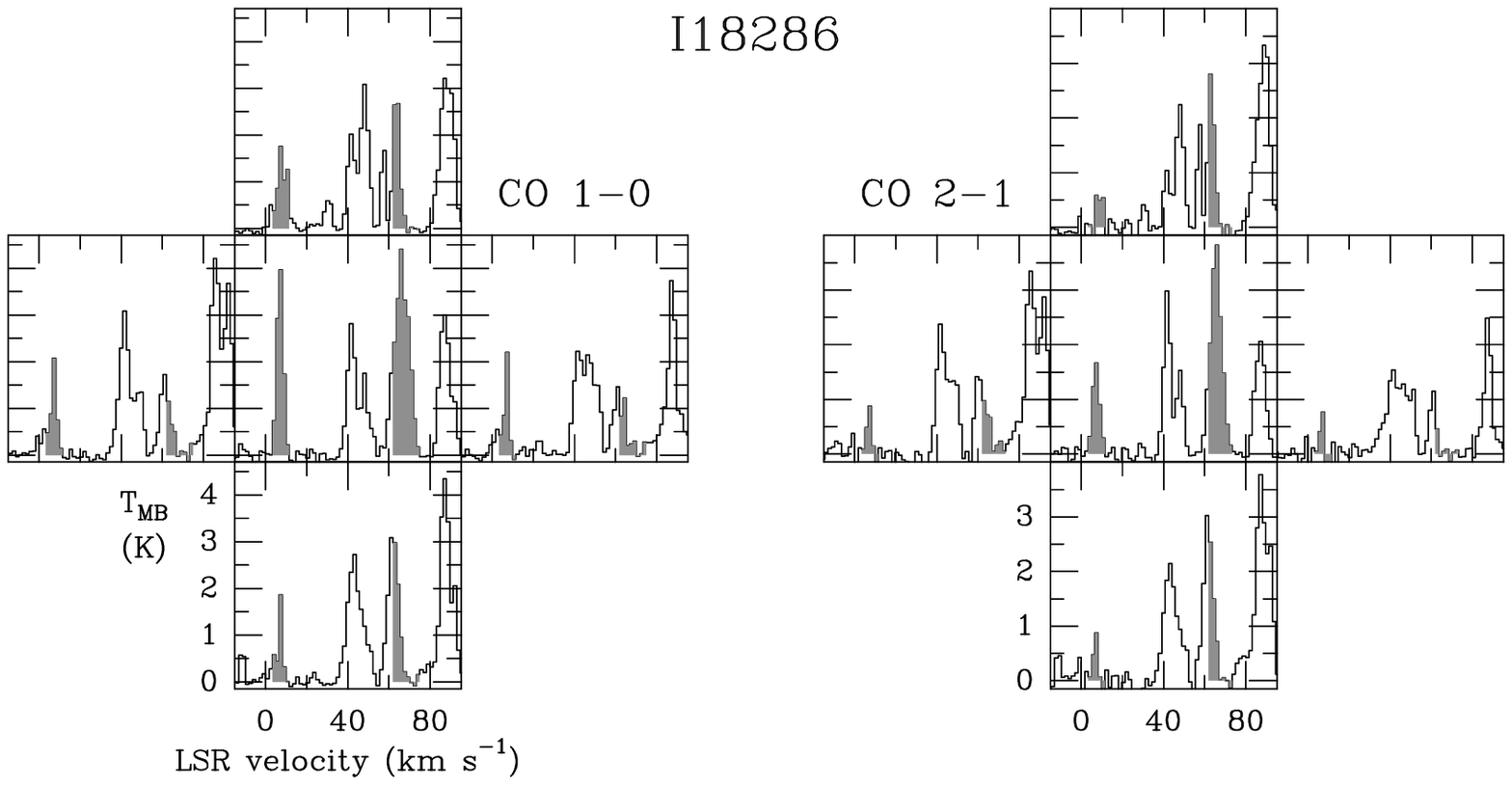}
      \caption{
Same as Figs.~5 and 6 for IRAS\,18286-959. In this case, a different pattern 
is proposed in the CO and $^{13}$CO associated with the WF. Two narrow velocity 
components, symmetrically located with respect to the stellar velocity, are found 
with a higher intensity towards the star position.
              }
   \end{figure*}
%

As was previously pointed out by \citet{ima09}, the CO spectrum of 
IRAS\,18286-0959 is rich in narrow components, due surely to its location in 
the Galactic plane. Nonetheless, two of these narrow components are 
particularly intense towards the star position and are weaker elsewhere. These 
two components lie at $\sim$ 7\kms\ and 66\kms, and are partly shaded in the 
Fig.~7 to facilitate comparison of the intensities among the five positions. 
The association of these narrow components to the WF is very tentative, but 
worth analyzing to the extent that these data allow \citep[see also the 
discussion by][]{ima09}.
 
Stellar velocities inferred from maser lines --due to their own nature and 
circumstellar origin-- may have large uncertainties of up to 20\kms\ 
\citep{tel89}. Even taking into account this potentially large uncertainty, the 
mean velocities of the OH and H$_2$O masers are probably not far from the 
stellar velocity, and the CO mean velocity is also close to it. Another 
important factor is that the line profile is contaminated by the contribution 
of foreground and background clouds, especially the blueshifted component.

\citet{ima09} (their Fig.~2c) detected three velocity components in the CO 
$J=3\rightarrow2$ line towards the star position, after subtracting the 
contribution of the off-source positions; the approximate velocities of these 
components are 10\kms, 40\kms, and 65\kms. In our data, the 40\kms\ component 
is not particularly intense at the star position, being noted only in the 
$J=2\rightarrow1$ line.

In the other two components, the observed (2-1)/(1-0) line ratio is lower than 
one. As we discuss below, this difference with respect to IRAS\,18460-0151 and 
IRAS\,18596+0315 is a signature of different physical and excitation 
conditions.

\section{Analysis and discussion}
\subsection{Possible origin of the detected CO}

To analyse and characterize the molecular gas associated with the three WFs 
that may show circumstellar CO emission, we isolated the intrinsic CO velocity 
components by subtracting synthesized off-source spectra (obtained, in turn, as 
the average of the four outer spectra of each source). The results are 
presented in Fig.~8 for the three sources and for all the observed lines.

%

   \begin{figure*}
   \centering
   \includegraphics[width=15cm]{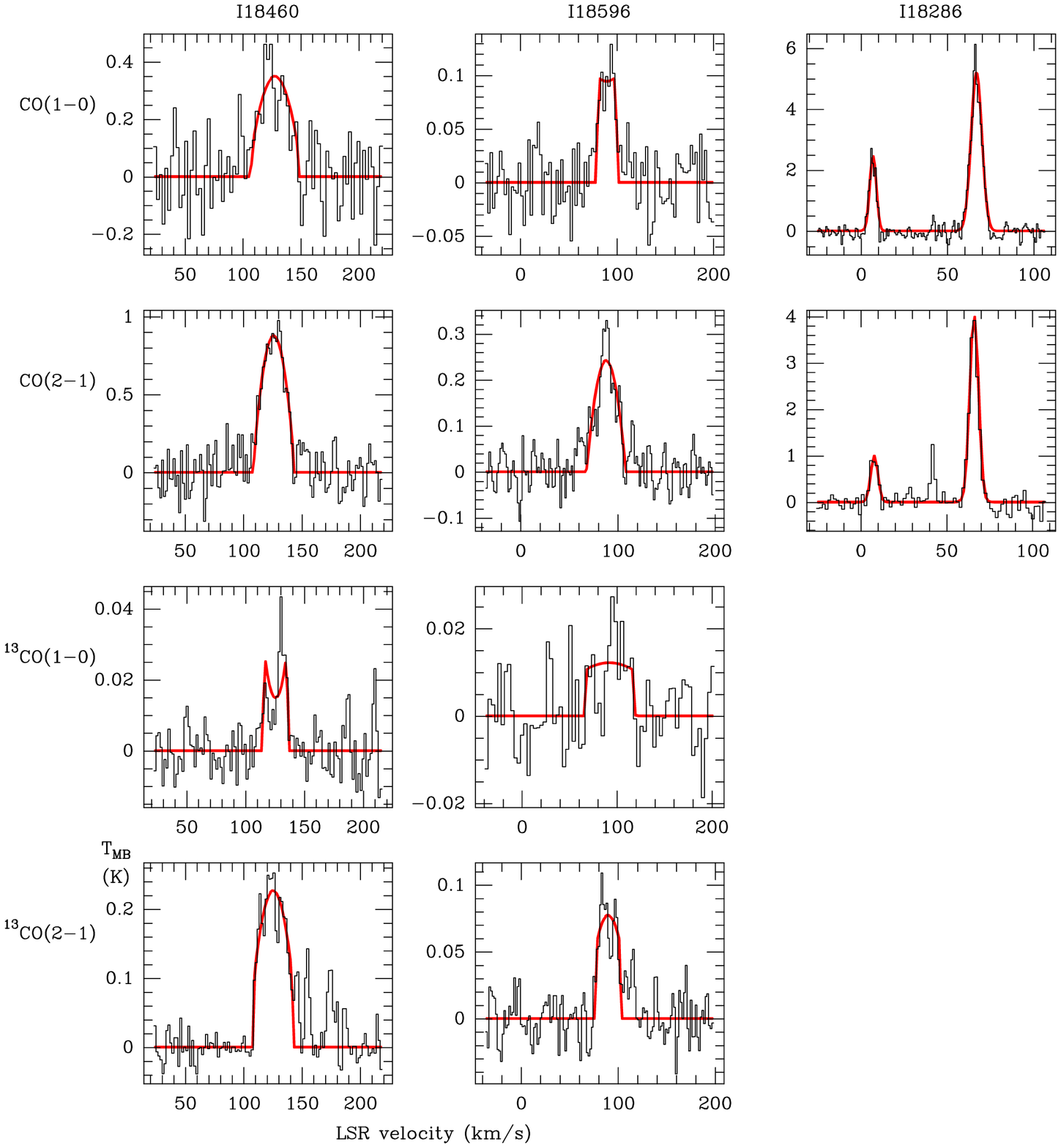}
   \caption{
Final spectra in the three cases reported as probable detections of CO and 
$^{13}$CO gas in WFs. These spectra are the result of subtracting synthesized 
off-source from on-source spectra for each source and transition. The 
superposed red lines are fits as circumstellar shells (for IRAS\,18460+0151 and 
IRAS\,18596+0315) or Gaussian profiles (for IRAS\,18286-0959). Each isotope and 
transition has been fitted independently (see text).
           }
   \end{figure*}
%

The wide component found in IRAS\,18460-0151 and IRAS\,18596+0315 strongly 
resembles those of the circumstellar envelopes found in different stages of 
late-type stars, such as planetary nebulae \citep[see, for instance,][]{buj05} 
and post-AGB stars \citep{buj01,tey06}. Wide components of 30--40\kms\ are 
commonlly associated with the previous AGB envelopes of these objects.

In the case of IRAS\,18286-0959 the velocity components are narrower than in 
the other two cases, and the emission (if it is associated with the source) 
probably arises from a smaller, more concentrated region; there are also some 
examples of circumstellar envelopes that show this double-peaked profile, 
although --to the best of our knowledge-- in C-rich evolved stars only 
\citep{yam93,buj04}. By increasing the degree of speculation, we can explore 
whether the CO arises from a thin dense shell that surrounds a bipolar cavity, 
as in the model of IRAS\,16342-3814 made by \citet{ima12}. The water jets 
recently studied by \citet{yun11} are organized in a double-helix structure, 
which may produce a pair of symmetric bullets, which are detected in CO.

The weaknesses of the lines (except IRAS\,18286-0959), the unresolved nature of 
the emission, and the lack of detailed information about the geometry and 
physics of the envelopes prevent a more detailed modelling. Nonetheless, 
first-order approximations were made by fitting the spectra. The shell method 
available in GILDAS was employed for IRAS\,18460-0151 and IRAS\,18596+0315. 
This method fits horn-type profiles for circumstellar envelopes, providing as 
outputs (1) the area of the spectrum; (2) $V_\mathrm{pk}$, the central 
velocity; (3) the expansion velocity ($V_\mathrm{exp}$), deduced from the full 
width at zero-level; and (5) $s_\tau$, a shape parameter that is measured as 
the horn-to-centre ratio, which depends on the line opacity and takes values 
from -1 (for optically thick lines, parabolic shape) to infinity (for 
double-peaked, optically thin lines); values close to zero are interpreted as 
optically thin, flat-topped lines. The method assumes uniform physical 
conditions and does not take into account possible asymmetries due to 
excitation or clumping. For IRAS\,18286-0959, this fitting method did not 
converge and we performed Gaussian fitting of the two components independently.

%
\begin{table*}
\caption{CO and $^{13}$CO parameters}
\centering                          
\begin{tabular}{lrrrrr}        
\hline\hline                 
\\
\multicolumn{6}{c}{SHELL FITTING} \\
\\
\multicolumn{1}{c}{Source} & \multicolumn{1}{c}{Line} & \multicolumn{1}{c}{Area} & 
\multicolumn{1}{c}{$V_\mathrm{pk}$} & \multicolumn{1}{c}{$V_\mathrm{exp}$} & 
\multicolumn{1}{c}{$s_\tau$}\\
& & \multicolumn{1}{c}{K\,\kms} & \multicolumn{1}{c}{\kms} & \multicolumn{1}{c}{\kms} \\
\hline
\\
IRAS\,18460-0151 & CO $1\rightarrow0$        & 10.6(2.6) & 127.1(3.5) & 19.9(0.9) & -0.72(0.21) \\
IRAS\,18460-0151 & CO $2\rightarrow1$        & 21.3(1.7) & 125.3(1.9) & 18.3(0.6) & -0.85(0.08) \\
IRAS\,18460-0151 & $^{13}$CO $1\rightarrow0$ &  0.4(0.1) & 125.0(5.1) & 10.3(3.0) &  1.02(0.71) \\
IRAS\,18460-0151 & $^{13}$CO $2\rightarrow1$ &  6.6(0.6) & 126.3(1.4) & 21.8(0.5) & -0.96(0.06) \\
\\
IRAS\,18596+0315 & CO $1\rightarrow0$        &  2.2(0.4) &  89.9(1.4) & 16.0(1.4) & -0.95(0.17) \\
IRAS\,18596+0315 & CO $2\rightarrow1$        &  6.5(0.9) &  87.3(1.8) & 18.8(1.8) & -0.83(0.54) \\
IRAS\,18596+0315 & $^{13}$CO $1\rightarrow0$ &  0.8(0.2) &  83.1(6.2) & 23.9(5.1) &  0.06(0.52) \\
IRAS\,18596+0315 & $^{13}$CO $2\rightarrow1$ &  2.0(0.3) &  90.3(1.6) & 18.3(1.5) & -0.93(0.27) \\
\\
\hline
\end{tabular}
\begin{tabular}{llrrrr}        
\\
\multicolumn{6}{c}{GAUSSIAN FITTING} \\
\\
\multicolumn{1}{c}{Source} & \multicolumn{1}{c}{Line} & \multicolumn{1}{c}{Area} & 
\multicolumn{1}{c}{$V_\mathrm{pk}$} & \multicolumn{1}{c}{$\Delta V$} & 
\multicolumn{1}{c}{$T_\mathrm{pk}$}\\
& & \multicolumn{1}{c}{K\,\kms} & \multicolumn{1}{c}{\kms} & \multicolumn{1}{c}{\kms}  & \multicolumn{1}{c}{K}\\
\hline
\\
IRAS\,18286-0959 & CO $1\rightarrow0$\,blue & 12.0(0.5) &  6.5(0.5) & 4.6(0.4) & 2.5(0.3) \\
IRAS\,18286-0959 & CO $1\rightarrow0$\,red  & 41.2(0.6) & 66.5(0.6) & 7.4(0.7) & 5.2(0.4) \\
\\
IRAS\,18286-0959 & CO $2\rightarrow1$\,blue &  5.3(0.6) &  7.5(0.5) & 4.9(0.6) & 1.0(0.2) \\
IRAS\,18286-0959 & CO $2\rightarrow1$\,red  & 26.1(0.7) & 65.6(0.4) & 6.1(0.6) & 4.0(0.3) \\
\\
\hline
\end{tabular}

\tablefoot{One-sigma errors within parenthesis.}
\end{table*}
%

Table 3 shows the results of the fitting, and the thick red lines of Fig.~8 
plot them superposed on the observed spectra. The fittings are sufficient as a 
first approximation. The mean velocities agree very well with the assumed 
systemic stellar velocity deduced from the masers. In IRAS\,18460-0151 and 
IRAS\,18596+0315, the two CO lines and the $J=2\rightarrow1$ line of $^{13}$CO 
are optically thick, while the $J=1\rightarrow0$ line of $^{13}$CO is 
optically thin in both sources. In any case, the fitting for the last line is 
the least reliable of all, particularly for IRAS\,18596+0315. Some hints of 
possible asymmetries are also noted and are worth to be studied in detail 
during follow-up observations. 

Average expansion velocities (weighted by the inverse square of the errors) are 
20.1\kms\  and 17.7\kms\ for IRAS\,18460-0151 and IRAS\,18596+0315. These 
velocities are consistent with an AGB origin of the molecular gas, as mentioned 
before.

From Table 3 we can also quantify some important line ratios. The CO and 
$^{13}$CO $(2-1)/(1-0)$ ratio varies between 2 and 3 in IRAS\,18460-0151 and 
IRAS\,18596+0315, except for $^{13}$CO in IRAS\,18460-0151; this unusually high 
line ratio is due to the low velocity-integrated temperature measured in the 
$^{13}$CO $J=1\rightarrow0$ line. For IRAS\,18286-0959, the CO $(2-1)/(1-0)$ 
ratios are 0.4 and 0.6 for the blue and red components; these values, more 
typical of diffuse clouds, indicate different excitation conditions in this 
source.

Another interesting reading of the Table 3 may be made by computing the 
CO/$^{13}$CO line ratio in IRAS\,18460-0151 and IRAS\,18596+0315 independently 
for each $J$ transition. In all cases except for the $J=1\rightarrow0$ case of 
IRAS\,18460-0151 these ratios are extremely low, between 2.6 and 3.3. It is 
worth noting that \citet{ima12} have measured similar values in 
IRAS\,16342-3814. This low line ratio may be interpreted in terms of high 
opacity of the CO line, or a real decrease of the isotopic ratio produced in 
specific events of the star evolution, such as the first dredge-up at the 
beginning of the RGB \citep{kar07}, or the deep mixing at the end of the 
same phase \citep{egg08}. This last hypothesis, first discussed by 
\citet{he08}, is supported by the fact that a low $^{12}$CO/$^{13}$CO line 
ratio is very common in PNe \citep{buj05} and proto-PNe \citep{buj01}.

The question arises why CO is detected in some WFs (IRAS\,18460-0151, 
IRAS\,18596+0315, and probably IRAS\,18286-0959), while in others (e.g., 
IRAS\,16552-3050, IRAS\,18113-2503, or IRAS\,19134+2131) it is clearly absent. 
In the remaining four sources (IRAS\,18043-2116, OH12.8-0.9, \object{W43A}, and 
IRAS\,19190+1102), Galactic contamination prevents any conclusive association 
or non-association of CO to the WFs, because in these cases there are intense 
spectral components in the stellar velocity range. Unfortunately, the few 
detections and unambiguous non-detections preclude any statistically 
significant discussion, and all suggestions must needs be highly speculative. 
We do not see any correlation of detectability with obvious physical 
parameters, such as the source distance or the envelope masses (estimated by 
Dur\'an-Rojas et al., in preparation). It is possible that CO is detected in 
WFs where the abundance of this molecule is enhanced. 

To explore this possibility, we investigated the morphology of the water maser 
emission where high angular resolution was available. In the sources without a 
clear detection of CO emission, the water masers trace only bipolar jets, with 
well-separated redshifted and blueshifted components 
\citep{ima07b,sua08,gom11}, while in IRAS\,18460-0151 there is a central maser 
structure tracing low-collimating mass loss. If shocks enhance CO abundance in 
the gas phase (due to chemical reactions or evaporation of ices), the detection 
of CO lines would be favoured in sources with current episodes of 
low-collimation mass-loss (in addition to the collimated jets), where shocked 
regions would occupy a large solid angle. However, the proper evaluation of 
this possibility would require interferometric observations, to determine 
whether CO emission is co-located with water masers.

\subsection{Physical parameters}

Aiming to shed more light on the origin of the detected CO and $^{13}$CO, and 
also to estimate the physical parameters of the envelopes, we have proceeded 
with non-LTE radiative modelling, using our own code of the {\it Large Velocity 
Gradient} (LVG) approach. For a given set of physical conditions --mainly the 
kinetic temperature \tk, H$_2$ volume density \n, line width and CO column 
density $N$(CO)--, this code iterates and computes the population of the 
rotational levels, and predicts some line intensities. More details of the LVG 
methodology is provided in the appendix. 

The code did not provide satisfactory results for IRAS\,18286-0959 for either 
of the two components. The low (2-1)/(1-0) line ratio that is observed could 
only be reproduced at \n\ values well below $10^2$\,cm$^{-3}$, typical of 
interstellar clouds, and $N$(CO) abnormally high. Nonetheless, and regarding 
our observational findings and those of \citet{ima09}, this case should not be 
ruled out because the CO is clearly contaminated by foreground or background 
Galactic emission.

For the other two sources, a summary of the best-fit results is shown in Table 
4. For IRAS\,18460-0151, the LVG approach does not provide any solution for 
\tk\ above 50\,K. For \tk=10\,K the solutions are optically thick, and for the 
range 20--50\,K the line emission is optically thin (or moderately thick for 
\tk=20\,K). The case of IRAS\,18596+0315, for a distance of 4.6\,kpc, is 
remarkably similar to IRAS\,18460-0151. In both cases, the total envelope 
masses are between $\sim$~0.2 and 0.4\,M$_{\odot}$, the mean densities around 
$10^{4}$\,cm$^{-3}$, and the mass-loss rates of the order of 
$10^{-4}$\,M$_{\odot}$\,yr$^{-1}$. The CO emission in IRAS\,18596+0315 is more 
opaque than in IRAS\,18460-0151. The $^{12}$C/$^{13}$C ratio --computed as the 
ratio between the column densities of $^{12}$CO and $^{13}$CO-- is low in both 
cases, although in IRAS\,18460-0151 it is not exceptionally low.

%
\begin{table*}
\caption{Results from the LVG modelling of IRAS\,18460-0151 and IRAS\,18596+0315}
\centering                          
\begin{tabular}{lccccccc}        
\hline\hline                 
\\
\multicolumn{8}{c}{MASS-RELATED PARAMETERS} \\
\\
\multicolumn{1}{c}{Source} & \multicolumn{1}{c}{\tk} & \multicolumn{1}{c}{$N$(CO)} & 
\multicolumn{1}{c}{mass} & \multicolumn{1}{c}{$n$(H$_2$)} & \multicolumn{1}{c}{$V_\mathrm{exp}$} 
& \multicolumn{1}{c}{$\dot{M}$} & \multicolumn{1}{c}{$^{12}$C/$^{13}$C}\\
\\
& \multicolumn{1}{c}{K} & \multicolumn{1}{c}{$10^{17}$\,cm$^{-2}$} & \multicolumn{1}{c}{M$_{\odot}$} 
& \multicolumn{1}{c}{$10^{3}$\,cm$^{-3}$} & \multicolumn{1}{c}{\kms} & \multicolumn{1}{c}{M$_{\odot}$\,yr$^{-1}$} \\
\hline
\\
IRAS\,18460-0151 &   10   & 1.57(0.44) & 0.20 & 7.85(2.4) & 20.1 & $2.6\,10^{-4}$ & 30 \\
IRAS\,18460-0151 & 20--50 & 1.50(0.32) & 0.19 & 7.50(2.1) & 20.1 & $2.5\,10^{-4}$ & 23 \\
\\
IRAS\,18596+0315$^{\rm a}$ &   10   & 2.70(0.83)  & 0.35 & 13.5(4.1) & 17.8 & $3.9\,10^{-4}$  &  9 \\
IRAS\,18596+0315$^{\rm a}$ & 20--50 & 1.91(0.26) & 0.25 & 9.55(2.3) & 17.8 & $2.8\,10^{-4}$  &  7 \\
\\
IRAS\,18596+0315$^{\rm b}$ & 30--50 & 18.0(4.8) & 2.32 & 90.0(23) & 17.8 & $2.6\,10^{-3}$ & 13 \\
\\
\hline
\end{tabular}
\begin{tabular}{lrrrrr}
\\
\multicolumn{6}{c}{OPACITIES} \\
\\
\multicolumn{1}{c}{Source} & \multicolumn{1}{c}{\tk\ (K)}& \multicolumn{1}{c}{CO 1--0} 
& \multicolumn{1}{c}{CO 2--1} & \multicolumn{1}{c}{$^{13}$CO 1--0}  & \multicolumn{1}{c}{$^{13}$CO 2--1} 
\\
\hline
\\
IRAS\,18460-0151 & 10 & 0.82(0.46) & 1.65(0.70) & 0.04(0.02) & 0.09(0.03) \\
IRAS\,18460-0151 & 20 & 0.23(0.10) & 0.90(0.60) & 0.01(0.01) & 0.06(0.04) \\
IRAS\,18460-0151 & 30 & 0.09(0.04) & 0.62(0.30) & \multicolumn{2}{c}{optically thin} \\
IRAS\,18460-0151 & 40 & 0.04(0.02) & 0.45(0.25) & \multicolumn{2}{c}{optically thin} \\
IRAS\,18460-0151 & 50 & 0.01(0.01) & 0.37(0.21) & \multicolumn{2}{c}{optically thin} \\
\\
IRAS\,18596+0315$^{\rm a}$ & 10 & 1.47(0.60) & 2.66(1.20) & 0.20(0.09) & 0.43(0.18) \\
IRAS\,18596+0315$^{\rm a}$ & 20 & 0.04(0.02) & 1.10(0.40) & 0.04(0.03) & 0.21(0.14) \\
IRAS\,18596+0315$^{\rm a}$ & 30 & 0.02(0.01) & 0.73(0.40) & 0.02(0.01) & 0.14(0.12) \\
IRAS\,18596+0315$^{\rm a}$ & 40 & \multicolumn{4}{c}{optically thin} \\
IRAS\,18596+0315$^{\rm a}$ & 50 & \multicolumn{4}{c}{optically thin} \\
\\
IRAS\,18596+0315$^{\rm b}$ & 30 & 0.97(0.40) & 2.90(1.20) & 0.10(0.03) & 0.32(0.07) \\
IRAS\,18596+0315$^{\rm b}$ & 40 & 0.63(0.35) & 2.12(0.70) & 0.06(0.02) & 0.22(0.09) \\
IRAS\,18596+0315$^{\rm b}$ & 50 & 0.38(0.20) & 1.40(0.50) & 0.04(0.01) & 0.14(0.07) \\
\\
\hline
\end{tabular}

\tablefoot{CO abundance of $10^{-4}$ assumed. Distances from Table 1. One-sigma errors within parenthesis.\\
$^{\rm a}$ Assumed distance is 4.6\,kpc.\\
$^{\rm b}$ Assumed distance is 8.8\,kpc.
}
\end{table*}
%

The line profiles (Fig.~8) and their corresponding shell fitting (Table 3) show 
that the envelope is characterized by moderate opacities. Taking this into 
account, the lower values of \tk\ should be favoured. Another strong argument 
that favours low \tk\ is provided after comparing the CO opacities of different 
transitions; the (2-1)/(1-0) opacity ratios are unacceptably high for the 
highest \tk, because they exceed the observed line ratios by far. The low 
values of \tk\ are consistent with a scenario in which CO arises from the outer 
parts of the envelope. 

Low temperatures also characterize the CO emission in most of the known 
post-AGB stars \citep[][and references therein]{buj01}. Furthermore, the values 
of the mass envelopes and the mass-loss rates agree very well with those 
derived previously \citep{kna85,buj01,buj05,deb10}. The high mass-loss rates 
derived from the LVG fitting are compatible with the mass-loss experienced by 
the AGB stars with the highest masses \citep{blo95}. Thus, the original masses 
of these sources are estimated to be in the range of 4--8 M$_\odot$.

The resulting parameters derived for the far kinematic distance of 
IRAS\,18596+0315 (8.8\,kpc) is, however, dramatically different with respect to 
the others. Firstly, the range of possible values of \tk\ is more restrictive, 
in the range 30--50\,K. Secondly, due to a higher $N$(CO), the mass grows to 
2\,M$_{\odot}$, \n\ is close to $10^{5}$\,cm$^{-3}$, and the mass-loss rate is 
of the order of $10^{-3}$\,M$_{\odot}$\,yr$^{-1}$. This case is indeed 
puzzling. The envelope mass is very large (although compatible with that 
obtained by Durán-Rojas et al., in preparation), and the mass-loss rate reaches 
a value that is too high within current models \citep{blo95}, and has not been 
observed in evolved stars with a low- and intermediate-mass progenitor. Unless 
something really unique occurs in this source, this result favour the nearest 
kinematic distance for this source, around 4.6\,kpc.

Finally, a word of caution is given with respect to the LVG estimates. We made 
strong assumptions about the size of the envelope and CO abundance. These 
assumptions and the uncertainty of the distances may produce variations of the 
parameters of a factor of two or even larger. The structure of CSEs in WFs is 
indeed more complex, due to the presence of shocks, bipolar outflows, or 
episodic mass-loss events. The complex heating and cooling processes due to 
excitation of molecules contribute to its complexity.

\section{Conclusions}

From this survey, we have discovered CO emission associated with two new WF 
sources: IRAS\,18460-0151 and IRAS\,18596+0315. A third case, IRAS\,18286-0959, 
is reported as tentative. These CO detections in WFs add new cases to 
IRAS\,16342-3814, the only one previously known \citep{he08}, and are the most 
promising candidates in which to study the molecular gas around WFs. 

The wide component found in IRAS\,18460-0151 was interpreted as the envelope of 
the former AGB stage. By non-LTE radiative modelling, we computed some physical 
parameters. The total envelope masses are around 0.2\,M$_{\odot}$, the 
mass-loss rates are of the order of $10^{-4}$\,M$_{\odot}$\,yr$^{-1}$, and the 
CO emission is moderately thick. The kinetic temperatures derived are rather 
low, in the range from 10\,K to 50\,K. Taking into account the moderate 
opacities derived from the line profiles, we favour the low values of \tk, 
which is also consistent with a scenario where the CO arises from the cold, 
outer parts of the envelopes.

For IRAS\,18596+0315 we also discovered a wide velocity component. For the near 
kinematic distance (4.6\,kpc), the LVG results were similar to those of 
IRAS\,18460-0151. For the far kinematic distance (8.8\,kpc), however, we 
predict higher kinetic temperatures (30--50\,K), large envelope masses 
($\simeq 2$\,M$_{\odot}$), and mass-loss rates one order of magnitude higher. 
These last values are hard to explain with the usual parameters found in other 
AGB envelopes, or with theoretical considerations \citep{blo95}.

For IRAS\,18286-0959, we found two narrow velocity components, symmetrically 
located with respect to the stellar velocity, $\sim30$\,\kms apart. Previous 
observations of the same source at the $J=3\rightarrow2$ transition 
\citep{ima09} failed to detect these components, but reported another narrow 
component close to the stellar velocity. Line contamination by Galactic 
background affects the computation of parameters and prevents further 
conclusions. Nonetheless, it is possible that the CO detected arises from 
somewhere in the jets traced by the water maser at 22\,GHz, and this case 
deserve a more detailed analysis in the future. We speculate whether CO is 
detected in WFs in which low-collimated mass-loss enhances the CO abundance in 
a relatively large area of the envelope.

The detection and study of circumstellar envelopes around WFs are the key to 
understanding how the mass ejections took place during the last stages of 
evolution of these objects. The high mass-loss rates derived from our CO data 
indicate that we are seeing the mass ejected at the end of the AGB phase, and 
that these objects derive from relatively massive (4--8\,M$_\odot$) 
progenitors. Sensitive, high-resolution observations are the natural follow-up 
of this work, and they may show how mass is ejected at the very end of the AGB 
phase, and its influence in shaping multipolar planetary nebulae.

\begin{acknowledgements}
JRR acknowledges support from MICINN (Spain) grants CSD2009-00038, 
AYA2009-07304, and AYA2012-32032. JFG, MO, OS and CD-R acknowledge support from 
MICINN grants AYA2008-06189-C03-01 and AYA2011-30228-C03-01, co-funded with 
FEDER funds. LFM acknowledges support from MICINN grant AYA2011-30228-C03-01, 
also co-funded with FEDER funds. JFG and MO are also supported by Junta de 
Andaluc\'{\i}a. The authors wish to thank Pico Veleta's staff for their kind 
and professional support during the observations. The careful reading and 
useful comments of the anonymous referee are also acknowledged, which certainly 
improved the paper.
\end{acknowledgements}


\begin{appendix}
\section{LVG modelling}

The LVG approach is a method that solves for different geometries the radiative 
transfer equations and the level populations iteratively and without assuming 
local thermodynamical equilibrium. It is based on the escape probability method 
first introduced by \citet{sob60}. This methodology has been and is widely used 
for determining the physical parameters of regions observed through their 
molecular rotational lines. An increasingly popular online version of this 
method is RADEX \citep{van07}, which we also used in this work for 
cross-checking.

The methodology consists of decoupling the radiative transfer from the level 
population trough introducing a parameter $\beta$ that measures the probability 
of a photon to escape the cloud. For an uniform sphere, it is computed by

\begin{equation}
\beta = (1-e^{-\tau})/\tau\,,
\end{equation}

\noindent where $\tau$ is the optical depth of a given transition. For an 
optically thin cloud, $\tau\rightarrow0$ and therefore $\beta\rightarrow1$, and 
all the photons escape the cloud. In contrast, $\beta$ becomes lower for an 
optically thick cloud, with a limit value of zero.

We modelled a uniform spherical cloud with an arbitrary radius 
$r_\mathrm{out} =10^{4}$\,AU, typical of AGB envelopes. A sphere of this size, 
located at the distances quoted in Table 1, would subtend at most some arcsec 
in the sky at the observed frequencies; therefore, we ran the code for a single 
point that concentrated all the emission from the envelope. The crude geometry 
that we had to introduce is, however, sufficient to provide first-order 
estimates of some global parameters, and is useful to constrain the theoretical 
models under development (Dur\'an-Rojas et al., in preparation).

We corrected the observed line intensities by the beam-filling factor (which is 
different for each source and frequency). A CO abundance with respect to H$_2$ 
($X_\mathrm{CO}$) of $10^{-4}$ and a H$_2$ mass abundance of 90\% were assumed. 
The collision partners are H$_2$ and He, and the collisional coefficients used 
were taken from \citet{gre76}. We note that, although reasonable, these 
assumptions certainly constitute an oversimplification and would translate into 
significant uncertainties in our results. The uncertainties quoted in Table 4 
(one-sigma) for $N($CO$)$ and \n\ are the result of the LVG model, considering 
only the uncertainties in the fitting (Table 3). For the total mass and the 
mass-loss rate, the uncertainties were computed by the standard formulae of 
propagation errors.

We ran the LVG code independently for the CO and $^{13}$CO species and for 
different values of \tk\ (from 10\,K to 100\,K, in steps of 10 K). For each 
temperature, we computed the radiative transfer for a grid of \n\ and 
$N($CO$)$, trying to find the best fit to the observed line intensities and the 
(2-1)/(1-0) line ratios. After determining the column densities, the opacities 
and envelope masses were obtained. Fig.~A.1 shows an example of the fitting, 
where the contours represent the observed $J=1\rightarrow0$ line intensity and 
the (2-1)/(1-0) line ratio; the position where the two sets of contours 
intersect determines the most probable value of $N($CO$)$ and \n\ for this 
value of \tk. The uncertainties in $N($CO$)$ and \n\ are also obtained from 
this fitting.

%

   \begin{figure}
   \centering
   \includegraphics[width=\columnwidth]{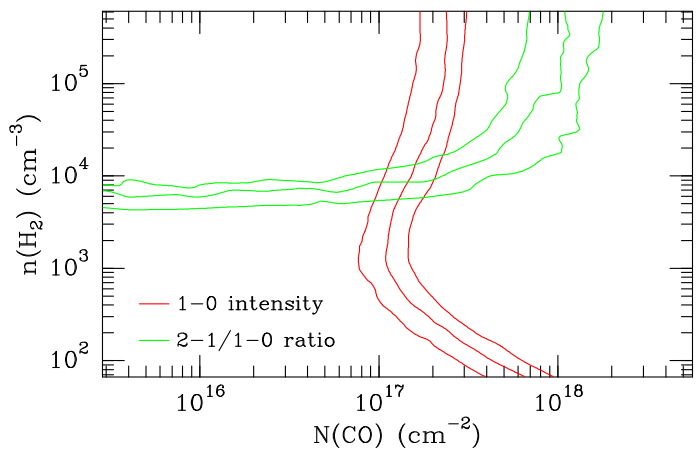}
   \caption{
Example of the fitting using the LVG approach. The contours correspond to 
the observed values of the $J=1\rightarrow0$ line intensity (red) and the 
(2-1)/(1-0) line ratio (green), together with their corresponding errors.
           }
   \end{figure}
%

For IRAS\,18596+0315, the procedure was applied twice, once for each of the 
possible kinematic distances (Table 1); for IRAS\,18286-0959, we also ran the 
code twice, once for each velocity component.

The mass was computed by integrating $N($CO$)$ over the projected disk: 

\begin{equation}
M=1.1\ (2\ \mathrm{m_H}) \ X_\mathrm{CO}^{-1}\ [\pi\ r_\mathrm{out}^2\ N(\mathrm{CO})],
\end{equation}

\noindent where m$_\mathrm{H}$ is the mass of the hydrogen atom, and the factor 
1.1 accounts for the mentioned H$_2$ mass abundance. 

Finally, the mass-loss rate was computed directly from the derived \n, with the only 
assumption of a stationary mass loss, and following the equation of continuity:

\begin{equation}
\dot{M}=4\ \pi\ r_\mathrm{out}^2\ V_\mathrm{exp}\ [1.1\ n(\mathrm{H}_2)],
\end{equation}

We remark that the mass-loss rate computed here is not the current mass-loss of 
the sources (that are probably in the post-AGB phase), but are the mass lost in 
the former AGB phase, during which the envelope was expelled.

\end{appendix}

\end{document}